\documentclass[12pt,preprint]{aastex}
\usepackage{graphicx}
\usepackage{amsmath}

\shorttitle{Reflection of waves generated in reconnection}
\shortauthors{Provornikova et al.}

\begin{document}

\title{Reflection of fast magnetosonic waves near magnetic reconnection region}

\author{E. Provornikova}
\affil{George Mason University, Fairfax, VA, 22030, USA}
\affil{Naval Research Laboratory, Washington, DC 20375, USA}

\author{J.M. Laming}
\affil{Naval Research Laboratory, Washington, D C 20375, USA}

\and

\author{V.S. Lukin }
\affil{National Science Foundation, Alexandria, VA, 22314, USA \footnote{Any opinion, findings, and conclusions or recommendations expressed in this material are those of the authors and do not necessarily reflect the views of the National Science Foundation.}}

\begin{abstract}
Magnetic reconnection in the solar corona is thought to be unstable to the
formation of multiple interacting plasmoids, and previous studies have shown
that plasmoid dynamics can trigger MHD waves of different modes propagating
outward from the reconnection site. 
However, variations in plasma parameters and magnetic field strength in the vicinity of a coronal reconnection site may lead to wave reflection and mode conversion.
In this paper we investigate the reflection and
refraction of fast magnetoacoustic waves near a reconnection site. Under a justified assumption of an analytically specified Alfv\'{e}n speed profile, we derive and solve analytically the full wave
equation governing propagation of fast mode waves in a non-uniform
background plasma without recourse to the small-wavelength approximation.
We show that the waves undergo reflection near the reconnection current sheet due to
the Alfv\'en speed gradient and that the reflection efficiently depends on the
plasma-$\beta$ parameter as well as on the wave frequency. In particular, we find that waves are reflected
more efficiently near reconnection sites in a low-$\beta$ plasma which is typical for the solar coronal conditions. Also, the
reflection is larger for lower frequency waves while high frequency waves
propagate outward from the reconnection region almost without the
reflection. We discuss the implications of efficient wave reflection near
magnetic reconnection sites in strongly magnetized coronal plasma for
particle acceleration, and also the effect this might have on First
Ionization Potential (FIP) fractionation by the ponderomotive force of
these waves in the chromosphere.
\end{abstract}

\keywords{Sun: corona -- Sun: magnetic fields -- magnetic reconnection -- magnetohydrodynamics (MHD)}

\section{Introduction}

In the context of the solar corona, magnetic reconnection and waves are often
studied separately despite the close relation between these two phenomena.
Magnetic reconnection can be a source of waves and waves can destabilize
magnetic null points (where $\mathbf{B}=0$) and trigger reconnection
processes \citep{mck09, lee14}. Presence of magnetic nulls can also lead to wave mode
conversion and greatly affect transfer of wave energy in the corona \citep{tarr17}. In
solar coronal plasma conditions with large Lundquist number (S), reconnection
current sheets are unstable to the secondary tearing instability with the
formation of complex dynamic structure with multiple plasmoids (flux ropes)
and X-points \citep{loureiro05, huang10, wyper14}. It is natural to expect
that such a dynamic process generates MHD waves of different modes (Alfv\'en
waves, fast and slow magnetosonic waves). The formation, growth, merging and
ejection of plasmoids can excite waves propagating outward from the
reconnection region. 
%Reconnection of magnetic field lines happens locally in
%a small region but causes a reconfiguration of magnetic field structure on
%larger scales on the order of the size of the solar active region.
%Reconfiguration perturbs the system and hence inevitably produces waves. In
%solar fares where magnetic reconnection is thought to be a key process
%magnetic energy is released impulsively. Such high energy impulsive events
%can drive waves and pulsations in solar plasma \citep{mcl17, liu11}. 
What
fraction of released magnetic energy in reconnection is transferred to wave
energy and what parameters determine this fraction are still not understood.

At present there are no direct observations confirming that magnetic
reconnection  drives waves in the solar corona. This is, in part, due to the
difficulty in observing lower emission intensities compared to the bright
emission from solar flares.
However some observations suggest the presence of waves and/or oscillations
that are due to magnetic reconnection in flares. \citet{brannon15} analyzed
observations from Interface Region Imaging Spectrograph (IRIS) of oscillating
flare ribbons in an M-class flare event. Flare ribbons appear as elongated
emission formed by the hot chromospheric plasma evaporated in response to the
energy deposit from coronal flare plasma. The structure and dynamics of the
ribbon emission is thought to serve as observational proxy for processes in
the flare reconnection current sheet. In the event, the ribbons  displayed
coherent substructure during the impulsive phase of the flare.
\citet{brannon15} proposed that the ribbon substructure is generated by
oscillations in flare loops which are driven by instabilities, most likely by
the tearing mode, of the reconnection current sheet. \citet{liu11} and
\citet{shen2012} reported arc-shaped quasi-periodic fast magnetoacoustic
waves propagating away from the flare site detected on Atmospheric Imaging
Assembly/Solar Dynamic Observatory (AIA/SDO). The periodicity of the fast
waves was found to be consistent with the periodicity of quasi periodic
pulsations in the flare light curve suggesting a common origin for these
oscillations. It is still unclear what processes determine flare pulsations
and what mechanisms excite propagating fast waves. The internal dynamics in
flare reconnection current sheet is a possible explanation of these
observations that needs to be investigated.

Several simulations demonstrated that plasmoid dynamics during magnetic
reconnection in the solar corona produces waves of different modes.
\citet{yang15} simulated interchange reconnection in the solar corona and
showed that the collision of the ejected plasmoids with the reconnection
outflow yields fast magnetoacoustic waves which propagate outward from the
reconnection region. Their simulation showed strong gradients of  Alfv\'en
speed across the magnetic field near the reconnection region where wave
reflection can happen. The merging of plasmoids in the reconnection current
sheet produces bigger plasmoids that can oscillate with a
period of tens of seconds \citep{jelinek17}. Such plasmoid oscillations are
another source of fast magnetoacoustic waves.  \citet{kig10} showed that
Alfv\'en waves and fast magnetoacoustic waves generated in reconnection may
carry a substantial part of released magnetic energy, more than $30\, \%$ for
Alfv\'en waves and $15\,\%$ for fast waves. The wave energy fluxes depend on
the inclination of reconnecting magnetic field and the plasma-$\beta$.

Waves produced in reconnection processes may play an important role in other
energetic processes associated with reconnection, for example particle
acceleration and plasma heating. In particular, waves are required for
scattering particles undergoing Fermi acceleration in reconnection current
sheets. In this process, similar to diffusive shock acceleration
\citep[DSA;][]{krymsky77}, particles scattered by waves move across the
reconnection current sheet, interact with flows incoming to the reconnection
site with the speed $V_{rec}$ and gain energy at each crossing increasing the
particle speed by $2V_{rec}$ \citep{drury12}. While in DSA fast
super-Alfv\'enic particles are required to excite waves that scatter
particles across the shock \citep[e.g.][]{melrose86,laming13}, in Fermi
acceleration during reconnection the waves can be produced by the reconnection
process itself, eliminating the necessity of initially super-Alfv\'enic ions.
In sufficiently compressed reconnection current sheets, the Fermi acceleration mechanism may also be able to produce hard energy spectra of suprathermal ions \citep{drury12}. Such high density current sheets with
compression ratio $\geq 4$ can form in magnetic nulls in the corona
\citep{prov16}.
The presence of suprathermal seed ion populations with hard energy spectra
is critical for injection into SEP acceleration at shocks in the low solar
corona \citep{laming13}.

Alfv\'en waves generated by reconnection in flares can propagate downward
along the flare loops and accelerate electrons in the legs of coronal loops
and at the chromospheric loop footpoints through several mechanisms
\citep{flhu08}.
%by the generated parallel electric fields, Fermi acceleration or in turbulence produced in a cascade of the waves to short wavelenghts.
Such a scenario could potentially help to explain the problem of a large
number of high energy electrons implied by the hard X-ray observations.
\citet{reep16} concluded that the dissipation of Alfv\'en waves in the upper
chromosphere causes heating very similar to the heating due to an electron
beam and leads to chromospheric evaporation. Despite of the potential importance of Alfv\'en waves in electron acceleration and chromosphere heating, their excitation process is not yet understood. Dynamic reconnection at the flare site presents one possible solution.

There is no question that the dynamic reconnection process produces waves.
Without focusing on how exactly waves are generated in reconnection we are
motivated by the question of how much wave energy produced by reconnection
can escape the reconnection site. In this paper we consider fast
magnetoacoustic waves only (leaving a study of Alfv\'en waves for a future
work). We examine how fast waves propagate in the vicinity of the
reconnection current sheet where plasma density and magnetic field strength
and therefore Alfv\'en speed are not uniform. We solve the full wave equation
for various wave frequencies and plasma-$\beta$ parameters analytically
without recourse to the small wavelength approximation. We show that due to
the Alfv\'en speed gradient, waves produced by reconnection dynamics
experience reflection. Waves with lower frequencies reflect more efficiently, with reflection efficiency further enhanced in $\beta \ll 1$ plasmas. We outline the assumptions of the analytical model in Section \ref{assumptions}. In Section
\ref{frequency}, we describe the coronal plasma parameters and the range of
wave frequencies considered in the calculations. In Section \ref{model} we
obtain an analytical solution of the wave equation for fast-mode waves
originating at the current sheet and propagating outward in the non-uniform
background plasma. We begin the section with a brief summary of the study by
\citet{hollweg84} of the propagation of Alfv\'en waves in an atmosphere with
exponential profile of plasma density. We discuss wave reflection in Section
\ref{reflection}. Implications of our results for particle acceleration in
reconnection and FIP effect are discussed in Sections
\ref{reflectionforparticle} and \ref{fip}, respectively. Conclusions are
presented in Section \ref{conclusions}.

\section{Assumptions in the model}\label{assumptions}

\begin{figure}
\includegraphics[scale=0.7]{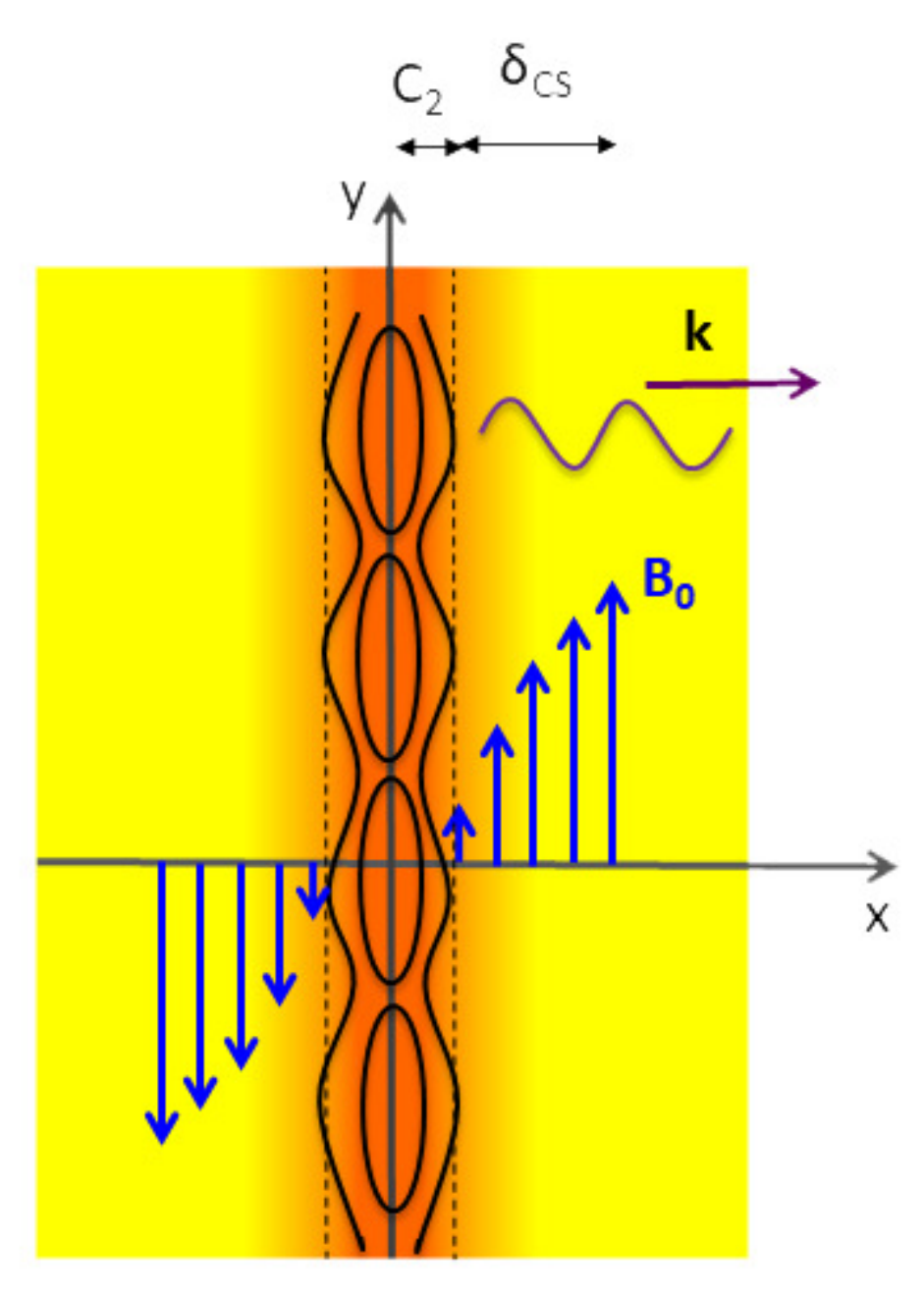}
\caption{ Schematic picture of generation of waves by the unstable reconnection region. Waves propagate outward from the reconnection region in plasma with density and magnetic field gradients (density decreases with distance from the center of the current sheet). $C_2$ is the half-width of the plasmoid-dominated current sheet, $\delta_{CS}$ is the density (magnetic field) gradient scale.
\label{cartoon}}
\end{figure}

We assume the following current sheet geometry: the reconnecting magnetic field $B_0$ is along the $y$-direction, and the current flow is along the $z$-direction.
Figure
\ref{cartoon} shows a schematic picture of waves propagating outward from the
plasmoid-dominated reconnection current sheet in a varying plasma background.
The current sheet is formed in strictly anti-parallel magnetic field (the
guide field is zero). The wave phase velocity is in the $x$-direction perpendicular to the B
field.

The simplifying assumptions in the analytical model are as follows: 1) For perturbations
and background quantities only $\partial/\partial x \neq 0$. 2) We consider fast magnetoacoustic waves
propagating in the direction normal to the current sheet, e.g. wave vector components are
$k_x \neq 0;\, k_y=0$. This is a reasonable assumptions for waves with $k_x >>
k_y$ generated by elongated plasmoids with high length to width ratio.
Further, since waves with significant $k_y$ would be refracted back to the current
sheet, we only consider those waves that otherwise could
escape. 4) The amplitudes of waves are small so that MHD equations can
be linearized. 5) In the linearization of the MHD equations, we assume that
the velocity of the background flow (inflow to the reconnection site
$V_{in}$) is zero. While in reconnection regions the velocity of the inflow
is not zero, this approximation is reasonable since the inflow is expected to be
significantly sub-Alfvenic $V_{in} \sim 0.01 V_A$ \citep{huang10, uzdensky10} and fast mode waves propagate with
the velocity $c_f=\sqrt{V_A^2 + c_s^2} \sim V_A$ in a strongly magnetized
plasma. Here $c_s$ is the sound speed in plasma.
6) Wave damping is neglected.
7) The background plasma is isothermal $T_i=T_e =T_0$. This implies that the sound speed is constant which simplifies the analytical treatment of the problem.
Note that we will find a solution to the full wave equation without using the WKB approximation when a small wavelength is assumed. We will be considering waves of various wavelengths including those of the order of the half-thickness of the current sheet $\delta_{CS}$, where  $\delta_{CS}$  is defined as the thickness of the region where plasma parameters change from their values in the current sheet to the values in the surrounding plasma undisturbed by reconnection (see Figure
\ref{cartoon}).

Table 1 presents a set of parameters of the coronal plasma for different
plasma-$\beta$ and characteristic temporal and spatial scales used in our
model.
 %The parameters are characterisctic for the quiet regions in solar coronal.
We choose the characteristic half-thickness of the current sheet to be $10^4$
km which is in the range of values inferred from observations $10^3 - 10^5$
km \citep{lin15, savage10, liu10}. To investigate wave reflection in current
sheets in coronal plasma with different conditions we will vary
plasma-$\beta$ in the range $0.02<\beta<1$.

\begin{table}
\begin{center}
\caption{Characteristic parameters of plasma and magnetic field in the solar corona.\label{tbl-1}}
\begin{tabular}{crrrrrrrrrrr}
\tableline\tableline
Parameter & Cool corona & Hot corona  \\
\tableline
Electron density $N_0$, $cm^{-3}$ & $10^{9}$ & $10^{9}$  \\
Temperature $T_0$, K & $10^6$ &  $3 \times 10^6$    \\
Magnetic field $B_0$, G & 10 & 10  \\
Plasma-$\beta$ parameter $\beta = p_{th}/p_{mag}$ & 0.07 & 0.2  \\
Alfv\'en speed $V_A$, km/s & 690 & 690  \\
Current sheet thickness $\delta_{CS}$, km & $10^4$ & $10^4$ \\
Characteristic timescale $\tau_A = \delta_{CS}/V_A $, s & 14.5 & 14.5 \\
Characteristic frequency  $\nu^*$, $s^{-1}$ & 0.07 & 0.07 \\
\tableline
\end{tabular}
\end{center}
\end{table}

\section{Frequencies of fast-mode waves}\label{frequency}

We consider the propagation of fast magnetoacoustic waves that are produced by the complex unstable magnetic reconnection process dominated by multiple plasmoids. Simulation results of \citet{yang15} support the idea that plasmoid ejections from the X-point generate fast waves propagating outward from the reconnection site. Thus it is reasonable to assume that frequencies and spatial scales of generated waves are related to those of the plasmoid dynamics in the reconnection current sheet. We will consider waves in the frequency range
\begin{eqnarray}
\nu_{plsm} < \nu << \nu_{i}
\end{eqnarray}
where $\nu_{plsm}$ is the minimal frequency of plasmoid ejection in the current
sheet and $\nu_{i}$ is the ion-ion collision frequency in plasma. Wave
frequencies have to be much smaller than the collision frequency since we
describe the plasma as a fluid. For the solar coronal plasma with
characteristic parameters in Table 1 the typical ion collision frequency is
around $\nu_{i} \sim 2 \, s^{-1}$.  The frequency of plasmoid ejection can be
estimated taking a ratio of the upstream Alfven speed and the characteristic
length of the plasmoid $\nu_{plsm}=V_A/\lambda_{plsm}$.

It is reasonable to accept that a nonlinearly formed plasmoid is always longer than the thickness of the current sheet. \citet{loureiro12} presented resistive MHD simulations of reconnection at high Lundquist numbers up to $10^7$ showing formation of multiple elongated plasmoids with the length exceeding plasmoids width and much larger than the thickness of the current sheet between plasmoids. We assume a typical half-thickness of the current sheet to be $\delta_{CS}=10^4$ km and a plasmoid half-length that is few times larger $5 \times 10^4$ km.
%Observations of a flare by Takasao et al. (2012) show plasma blobs interpreted as plasmoids formed during reconnection with size in range $2-4\,\, \times 10^3$ km.
Taking this estimate we obtain a range of frequencies and corresponding wave periods to
consider;
\begin{eqnarray}
0.01\, s^ {-1} < \nu << 2\, {\rm s}^{-1}\\
0.5\, s << T < 2 \, {\rm min}.
\end{eqnarray}
This frequency range is consistent with wave frequencies produced
by plasmoid dynamics in previous MHD simulations. \citet{yang15} obtained
frequencies of fast waves below $0.25 \,s^{-1}$. \citet{jelinek17} reported
wave period $\sim 25\,s$ generated by oscillating plasmoid. Below we will
compare results for different parameters in dimensionless units therefore for
the reference our frequency range in dimensionless units is $0.2 < \hat{\nu} <<
30$.

\section{Analytical model}\label{model}

In this section we derive and solve the wave equation for fast mode waves
propagating outward from the reconnection site in a non-uniform background
plasma. In our analysis we will refer to the study of propagation of Alfv\'en
waves in the atmosphere with the exponential profile of plasma density by
\citet{hollweg84}. In the next subsection we briefly summarize this study.

 \subsection{Propagation of Alfv\'en waves in a non-uniform plasma}

Hollweg (1984) considered a propagation of small-amplitude transverse (and
non-compressive) Alfv\'en waves along an untwisted magnetic field
$\mathbf{B}_0$ on a static background ($\mathbf{V}_0 = 0$). All quantities
are axisymmetric relative to the vertical $z$-axis of asymmetry directed
along $\mathbf{B}_0$ field implying that $\partial/\partial \theta =0$ where
$\theta$ is the azimuthal angle.

Near the axis of asymmetry the wave equation for Alfv\'en waves has the form
\begin{eqnarray}
\frac{\partial^2 x}{\partial t^2}=V^2_A \frac{\partial^2 x}{\partial s^2}. \label{weq}
\end{eqnarray}
Here $x \equiv \delta v_{\theta}/r$, $ \delta v_{\theta}$ is the velocity
disturbance and $s$ is the distance along the magnetic field line. When the
Alfv\'en speed $V_A$ varies exponentially $V_A \varpropto e^{s/2h}$, then the
equation has the following solution in terms of Hankel functions $H_0^{(1)}$
and $H_0^{(2)}$ of the first and second kinds, respectively,
\begin{eqnarray}
x = [aH_0^{(1)}(\xi)+b H_0^{(2)}(\xi)] e^{i\omega t}, \label{dv}
\end{eqnarray}
where $\xi \equiv 2 h \omega/V_A(s)$, $\omega = 2 \pi \nu$, angular frequency, and $a$ and $b$ are complex constants.
The corresponding magnetic field fluctuation can be derived from the
linearized induction equation,
\begin{eqnarray}
\delta \mathbf{B}_{\theta} = - \frac{i r \mathbf{B}_{0s}}{V_A}[aH_1^{(1)}(\xi)+b H_0^{(2)}(\xi)] e^{i\omega t}. \label{dB}
\end{eqnarray}
The time-averaged Poynting flux, $<\mathbf{S}>$, of the wave is
\begin{eqnarray}
<\mathbf{S}> = - \mathbf{B}_0 <\delta v_{\theta} \delta B_{\theta}> /4 \pi .
\end{eqnarray}
With the solution for $\delta v_{\theta}$ and $\delta B_{\theta}$ given  by
(\ref{dv}) and (\ref{dB}) the Poynting flux along $\mathbf{B}_0$ is
 \begin{eqnarray}
<S_s> = \frac{B^2_{0s}r^2}{8 \pi^2 h \omega} \left( |a|^2 - |b|^2 \right). \label{pflux}
\label{Pflux}
\end{eqnarray}
From the form of Eq. (\ref{pflux}), the parts of Eq. \ref{dv}
associated with $H_0^{(1)}$ and $H_0^{(2)}$ are identified as the
upward-propagating and downward-propagating waves, respectively.

Now consider a two-layer model in which $V_A$ varies exponentially for $s<0$
while $V_A = V_{Ac}= constant$ for $s>0$. $V_A$ and $B_{0s}$ are assumed to be
continuous at $s=0$. Suppose that there is some unspecified source of waves
in $s<0$. Above the source the solution will be given by equations (\ref{dv})
and (\ref{dB}) in $s<0$. But in $s>0$ the solution to equation (\ref{weq})
must be a pure plane wave
\begin{eqnarray}
x = c e^{(i \omega t - i k s)}\\
\delta \mathbf{B}_{\theta} = - \frac{k \mathbf{B}_{0s}}{\omega} \delta v_{\theta},
\end{eqnarray}
where $k = \omega/ V_{Ac}$ is the wavenumber and c is now a complex constant.
At $s = 0$ the disturbances $\delta v_{\theta}$ and $\delta
\mathbf{B}_{\theta}$ are continuous. These boundary conditiions allow to
determine unknown complex constants. Then the wave energy reflection
coefficient, R, i.e. the ratio of downgoing energy flux to upgoing energy
flux in $s<0$ can be obtained as
\begin{eqnarray}
R = \frac{|b|^2}{|a|^2}. \label{refc}
\label{Href}
\end{eqnarray}
This completes the brief discussion of propagation and reflection of Alfv\'en
waves in the non-uniform  two-layer atmosphere presented in Hollweg (1984).
In the following subsection we will carry out similar analysis for the fast
mode waves propagating in plasma with varying density and magnetic field.

\subsection{Propagation of fast waves in the neighborhood of a current sheet}

\begin{figure}
\includegraphics[scale=0.65]{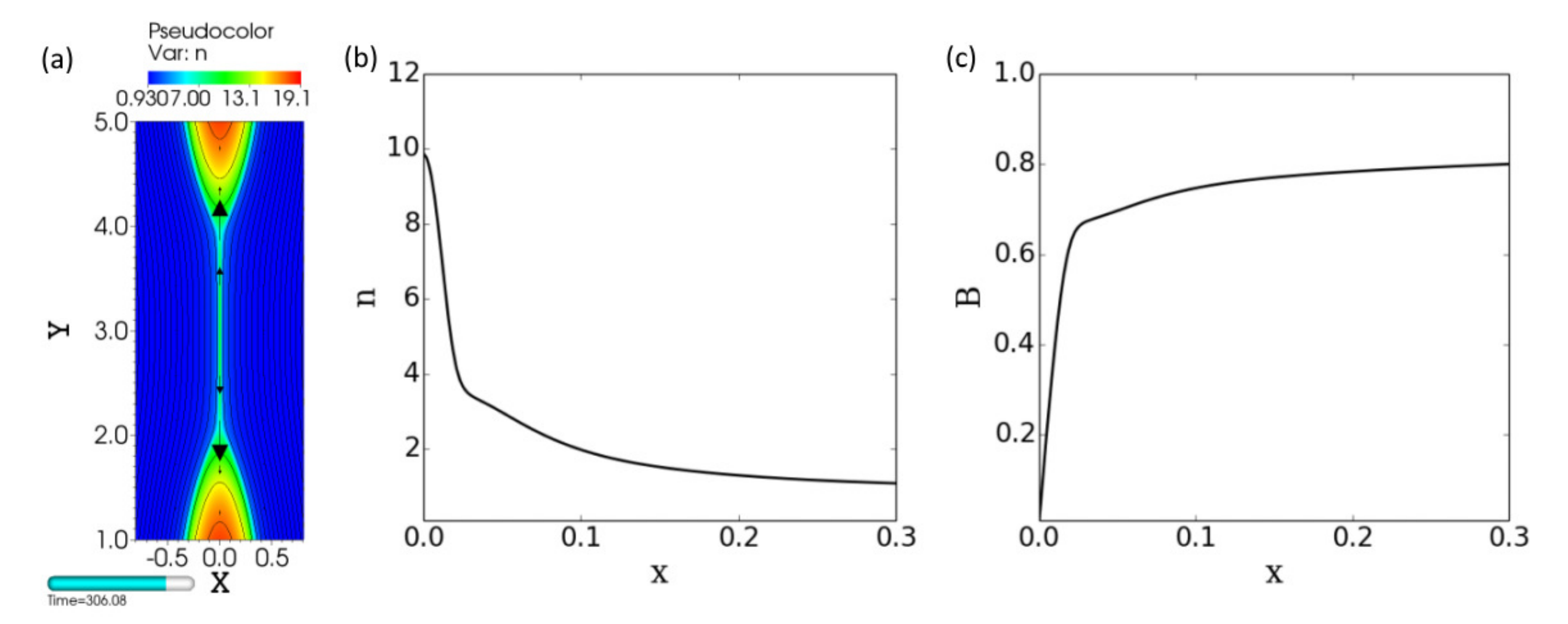}
\caption{ Resistive MHD simulation of reconnection in the isothermal Harris current sheet. a) colormap of plasma density $n$. Black arrows illustrate plasma velocity field. b) Density profile across the reconnection region at $x=0;\,y=3$. c) Magnitude of the reconnecting magnetic field across the reconnection region at $x=0;\,y=3$. This simulation is for plasma $\beta = 0.07$ and resistivity $\eta = 10^{-4}$ (or corresponding Lundquist number $S=10^4$).
\label{fig1}}
\end{figure}

\begin{figure}
\includegraphics[scale=0.53]{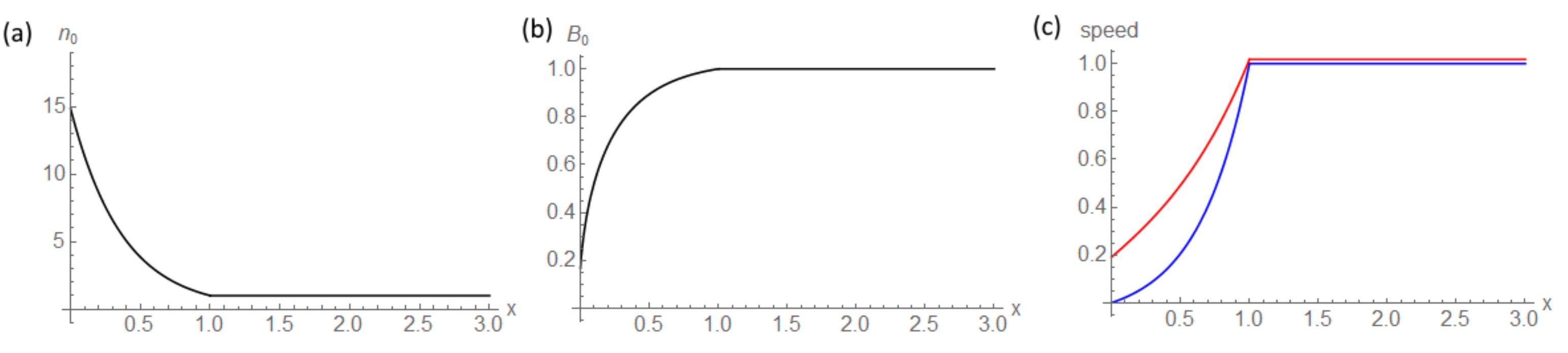}
\caption{ Typical profiles of plasma density (a) and reconnecting magnetic field component (b) used in the analytical model. Profiles of Afv\'en speed (blue curve) and fast magnetoacoustic speed (red curve) are shown on plot (c). Profiles were obtained assuming $\beta=0.07$ and $c_0=0.03$.
\label{fig2}}
\end{figure}

We choose to specify analytical profiles for the background plasma density and
magnetic field in the vicinity of the reconnection region. Reconnection
regions with strong plasma compressions are of particular interest since
these are potential sources of hard energy spectra of particles accelerated by the Fermi
process \citep{drury12, montag17}. In \citet{prov16} we performed resistive
MHD simulations of Sweet-Parker-like laminar magnetic reconnection in
different magnetic field geometries and with varying plasma parameters. In the highly
conductive coronal plasma the Lundquist number $S \sim 10^{14}$ and
reconnection current sheets are unstable to multiple plasmoid formation
(Loureiro et al. 2012). In Provornikova et al. (2016) we limited our
consideration to a single laminar reconnection region that , for the purposes of this paper, can represent the large scale reconnection region with the plasmoid substructure assumed to exist, and averaged over, within the macroscopic current sheet's diffusion layer.   Figure \ref{fig1} shows an example of
a simulated reconnection region with strong plasma compression, by a factor
of 5.
Panels b) and c) present the density and reconnecting field $B_y$ profiles
across the reconnection region (due to the symmetry only $x>0$ is shown). We
choose the analytical profiles that approximate profiles obtained in
simulations. Let us assume that away from the reconnection region the
undisturbed plasma is characterized by the plasma-$\beta$. The normalization parameters
are the number density $N^*$, the Alfv\'en speed $V^*=V_A = B^*/\sqrt{N^*
\mu_0}$, the magnetic field $B^*$, and the half-thickness of the current sheet
$\delta_{CS}$ (Table 1). We approximate the density profile with the
following piecewise function

\begin{eqnarray}
n(x) =  \left\{
\begin{array}{ll}
\left( \frac{1}{\beta } + 1\right) e^{-\frac{C_1x+C_2}{x0}}  & (0\leq x<1)\\
1 & (x \geq 1).\\
\end{array}
\right.
\label{nprofile}
%\nonumber
\end{eqnarray}
Here $n(x)$ is the normalized number density and $x_0=\frac{1}{ln(1/\beta
+1)}$. For the calculations we choose $C_2$ in the range $0<C_2<1$ and define
$C_1$ as $C_1 = 1 - C_2$. The constants are introduced to define the
background plasma variations in the region outside of the plasmoid dominated
current sheet with the half-width $C_2$ (see Figure \ref{cartoon}). Hereafter
parameters for region $0\leq x<1$ are denoted with the subscript 1 and for
region $x \geq 1$ with subscript 2. We assume a constant magnetic field in
region 2, $B(x) \equiv B_2=1$.
%In calculations we assume that $C_2<< 1$ and $C_1 = 1 - C_2$, namely $C_2=0.01$. Constants are chosen for the purpose to avoid %having infinite derivative of the $B$ profile at $x=0$. In other words we investigate propagation of waves starting from a very small distance %$\delta x =C_2=0.01$ from the center of the current sheet on Figure \ref{fig1}.
To obtain a profile for the magnetic field in region 1 we use the condition
of total pressure balance in the system which yields
%\begin{eqnarray}
%2 n_1 k_B T + \frac{B_1^2}{2 \mu_0} = const = 2 n_2 k_B T + \frac{B_2^2}{2 \mu_0} \nonumber
%\end{eqnarray}
% After normalization of pressure by ${B^*}^2/\mu_0$ and temperature by ${B^*}^2/(\mu_0 k_B N^*)$ we obtain;
%\begin{eqnarray}
%2 n_1 T + \frac{B_1^2}{2} =  \frac{\beta +1 }{2} \nonumber \\
%B^2= \beta+1 - \beta n \label{eqB2}
%\end{eqnarray}
%Here we used the fact that the plasma is isothermal and in region 2 the normalized temperature $T=\beta/4$. From equations (\ref{nprofile}) and (\ref{eqB2}) the magnetic field strength profile is given by:
\begin{eqnarray}
B(x) =  \left\{
\begin{array}{ll}
\sqrt{\left( 1+\beta \right) \left( 1 - e^{-\frac{C_1 x + C_2}{x0}} \right)}   & (0\leq x<1)\\
1 & (x \geq 1).\\
\end{array}
\right.
\label{bprofile}
\end{eqnarray}
The profiles of plasma density and magnetic field given by (\ref{nprofile})
and (\ref{bprofile}) are shown in Figure \ref{fig2} a) and b). The resulting
Alfv\'en speed and fast magnetoacoustic speed profiles are shown in Figure
\ref{fig2} c).

We derive a solution of the wave equation for region 1. Following the
standard procedure of linearization of a system of ideal MHD equations
\citep{priest14}, accounting for the non-uniform background we obtain an
equation for the velocity disturbance $\delta \mathbf{V'}$
%Then the system of linearized MHD equationsreduces to one equation for $\mathbf{V}_1$ by taking $\partial/\partial t$ of momentum and substitute $\frac{\partial \rho_1 }{\partial t}$, $\frac{\partial \rho_1 }{\partial t}$, $\frac{\partial \mathbf{B}_1 }{\partial t}$ from other equations (standard procedure):
\begin{eqnarray}
n_0 \frac{\partial^2 \delta \mathbf{V'} }{\partial t^2} =
\nabla \left( \left( \delta \mathbf{V'} \cdot \nabla \right) p_0 \right) +
\nabla \left( c_s^2 n_0 \nabla \cdot \delta \mathbf{V'} \right) +
\frac{1}{\mu_0} \left[ \nabla \times \left[ \nabla \times \left(\delta \mathbf{V'}
\times \mathbf{B}_0 \right) \right] \right] \times \mathbf{B}_0 \nonumber \\ +
\frac{1}{\mu_0} \left( \nabla \times \mathbf{B}_0 \right) \times \left[ \nabla
\times \left( \delta \mathbf{V'} \times \mathbf{B}_0  \right) \right], \label{v1eq}
\end{eqnarray}
where $n_0(x)$ and $\mathbf{B}_0(x)=(0,B_0(x),0)$ are the background number
density and magnetic field given by Eqs. (\ref{nprofile}) and
(\ref{bprofile}), respectively, $p_0$ is the plasma pressure, and $c_s$ is
the (uniform) sound speed $c_s^2 = 2 T= \beta/2$. We consider waves propagating along the
$x$-axis perpendicular to $\mathbf{B}_0$ and we
will look for a solution in a form $\delta \mathbf{V'} = (\delta V(x)  \exp
(-i \omega t), 0,0)$. Then equation (\ref{v1eq}) reduces to an equation for
$\delta V(x)$
\begin{eqnarray}
n_0 (c_s^2 + c_A^2) \frac{d^2 \delta V}{dx^2} +\frac{d \delta V}{d x} \left(  \frac{d n_0 (  c_s^2 + c_A^2 ) }{dx}  \right)  + \delta V n_0 \omega^2 =0
\label{eq3}
\end{eqnarray}
or
\begin{eqnarray}
\frac{d}{d x} \left( n_0 (  c_s^2 + c_A^2 ) \frac {d \delta V} {dx}  \right)  = - \delta V n_0 \omega^2 \label{final}
\label{eq21}
\end{eqnarray}
where $c_A$ is the non-uniform Alfv\'en speed. Substituting for $B_0$ and $n_0$ with the expressions
for region 1, Eqs. (\ref{bprofile}) and (\ref{nprofile}), we obtain
\begin{eqnarray}
n_0 (  c_s^2 + c_A^2 ) = n_0\beta /2 + B_0^2 = 1+\beta - \frac{1}{2} \left(1 + \beta \right)
e^{-\frac{C_1x + C_2}{x_0}}.
\end{eqnarray}
Equation (\ref{eq21}) becomes
\begin{eqnarray}
\frac{d}{d x} \left( \left(1+\beta - \frac{1}{2} \left(1 + \beta \right)
e^{-\frac{C_1x + C_2}{x_0}} \right) \frac {d \delta V} {dx}  \right)  =
 - \delta V \omega^2 \left( \frac{1}{\beta} +1 \right)e^{-\frac{C_1x + C_2}{x_0}}. \label{final}
\label{eq2}
\end{eqnarray}
Introducing the variable $\xi(x) \equiv e^{-\frac{C_1x + C_2}{x_0}}-1$, equation
(\ref{eq2}) reduces to
\begin{eqnarray}
\left(1 - \xi^2 \right)\frac{d^2 \delta V(\xi) }{d \xi^2} - 2\xi
\frac{d \delta V(\xi) }{d \xi} + \frac{2 x_0^2 \omega^2 }{\beta C_1^2}  \delta V(\xi) =0.
\label{eqxi}
\end{eqnarray}

The solution of this equation is a linear combination of the associated
Legendre functions of first and second kind, $P_p(\xi)$ and $Q_p(\xi)$
\citep[][p. 332]{abram65}. The order $p$ of the functions is found by solving
an equation $p(p+1)= \frac{2 x_0^2 \omega^2 }{\beta C_1^2}$ and taking the
positive root. The general solution of the equation (\ref{eqxi}) can be
written in the following form,
\begin{eqnarray}
\delta V(\xi) = c (P_p(\xi)-i \alpha Q_p(\xi))+d(P_p(\xi)+i \alpha Q_p(\xi)),
\label{sol}
\end{eqnarray}
%or
%\begin{eqnarray}
%\delta V(x) = C (P_p(e^{-x/x_0}-1)+IQ_p(e^{-x/x_0}-1))+D(P_p(e^{-x/x_0}-1)-IQ_p(e^{-x/x_0}-1))
%\end{eqnarray}
where the complex $c=c_r + ic_i$ and $d=d_r+id_i$ are coefficients to be
found, and $\alpha$ is a coefficient defined at $\xi =0$ as $\alpha =
\sqrt{-\frac{P_p(0) P_p'(0)}{Q_p(0) Q_p'(0)}}$ (see Appendix). Substituting
$\xi(x)$ one can obtain the expression for $\delta V(x)$. The solution
(\ref{sol}) multiplied by time dependence $e^{-i \omega t}$ represents the
velocity disturbance $\delta V'$ in a propagating fast-magnetoacoustic wave
in the non-uniform region 1.

The time-averaged energy flux $<F>$ of fast magnetoacoustic waves can be
calculated as a sum of the acoustic flux and Poynting flux and after
normalization has the form
\begin{eqnarray}
<F> = <\operatorname{Re} (\delta p') \, \operatorname{Re} (\delta V') + B_0
\operatorname{Re} (\delta V') \operatorname{Re} (\delta B')> = \\
=<c_s^2\operatorname{Re} (\delta \rho') \, \operatorname{Re} (\delta V') + B_0
\operatorname{Re} (\delta V') \operatorname{Re} (\delta B')>.
\nonumber
\label{flux}
\end{eqnarray}
Here $\delta B'=\delta B (x) e^{-iwt}$ and $\delta p'=\delta p (x) e^{-iwt} =
c_s^2 \delta \rho (x) e^{-iwt}$ are the $y$-components  of the magnetic field
disturbance and plasma pressure disturbance, respectively. From the
linearized induction equation $\frac{\partial \delta B'}{\partial t} +
\frac{\partial \left( \delta V' B_0 \right) }{\partial x} = 0$ one can obtain
$\delta B = -i/\omega(\partial(B_0 \delta V)/\partial x)$. Similarly from the
linearized continuity equation, $\delta \rho = -i/\omega(\partial(\rho_0
\delta V)/\partial x)$. Thus using (\ref{sol}) we can derive an expression
for the wave energy flux which has the form
\begin{eqnarray}
<F>=\frac{1}{2 \omega} \rho_0 c_f^2 \alpha (|c|^2-|d|^2)(P_p'Q_p - Q_p' P_p).
\label{flux1}
\end{eqnarray}
The expression for the energy flux of fast mode waves is similar to the
expression (\ref{Pflux}) derived by Hollweg (1984) and represents the
difference in the energy flux of outgoing waves and reflected waves. From the
form of Eq. (\ref{flux1}) we identify the parts of Eq. (\ref{sol}) as outgoing and
reflected waves, respectively,
\begin{eqnarray}
\delta V_{out} = c (P_p-i \alpha Q_p) e^{-i\omega t} \label{out}\\
\delta V_r = d (P_p+i \alpha Q_p) e^{-i\omega t}. \label{refl}
\end{eqnarray}
Now, similarly to Eq. (\ref{Href}) in the Hollweg study, we define the wave
energy reflection coefficient as the ratio of fluxes of outgoing and
reflected waves in Eq. (\ref{flux1}),
\begin{eqnarray}
R = \frac{|d|^2}{|c|^2}.
\label{refcoef}
\end{eqnarray}
To calculate the reflection coefficient for various wave frequencies and
plasma-$\beta$, it is required to determine complex constants $c$ and $d$.

In region 2, $x \geq 1$, where plasma parameters are constant the solution of
the equation (\ref{v1eq}) is a plane wave,
\begin{eqnarray}
\delta V_t = v_t e^{i(k_t x - \omega t)},\label{trnsm}
\end{eqnarray}
where $v_t$ is the amplitude of the wave, $k_t = \omega/c_f$ is the wave
number, $c_f$ is the (constant) fast-mode speed in region 2. The subscript
$t$ refers to the transmitted wave.

We have obtained a general solution for a fast magnetoacoustic wave
propagating in the region 1 with non-uniform background density and magnetic
field and a plane wave solution in uniform region 2. Since background $n_0$
and $B_0$, and therefore Alfv\'en speed $c_A$, are continuous at $x=1$
(see Figure \ref{fig2}), $\delta V$, $\delta \rho$ and $\delta B$ must also
be continuous. Thus, assuming that at $x=0$ there is some source of waves
with angular frequency $\omega$ and arbitrary small amplitude $V_b$, the
boundary conditions are as follows,
\begin{eqnarray}
\begin{array}{ll}
x=0:  & |\delta V_{out}| = V_b \,\, \text{and phase of} \,\, V_{out} \,\, \text{is zero} \\
x=1: & \delta V_{out} + \delta V_{r} = v_{t}  e^{i(\omega/c_f + \phi)},  \\  & \delta \rho_{out}
+ \delta \rho_{r} = \frac{\rho_0(1)}{c_f} v_{t} e^{i(\omega/c_f + \phi)}\\ & (\delta B_{out}
+ \delta B_{r})e^{\phi_1} = \frac{B_0(1)}{c_f} v_{t} e^{i(\omega/c_f + \phi)}.
\end{array}
\label{bc}
\end{eqnarray}
The conditions define the in-phase relation between $\delta V$, $\delta \rho$
and $\delta B$ in the transmitted wave since it is a fast wave propagating in
the uniform plasma. The phase $\phi_1$ is introduced to ensure that $\delta
B$ is continuous at $x=1$.

\section{Reflection of fast mode waves}\label{reflection}

\begin{figure}
\includegraphics[scale=0.5]{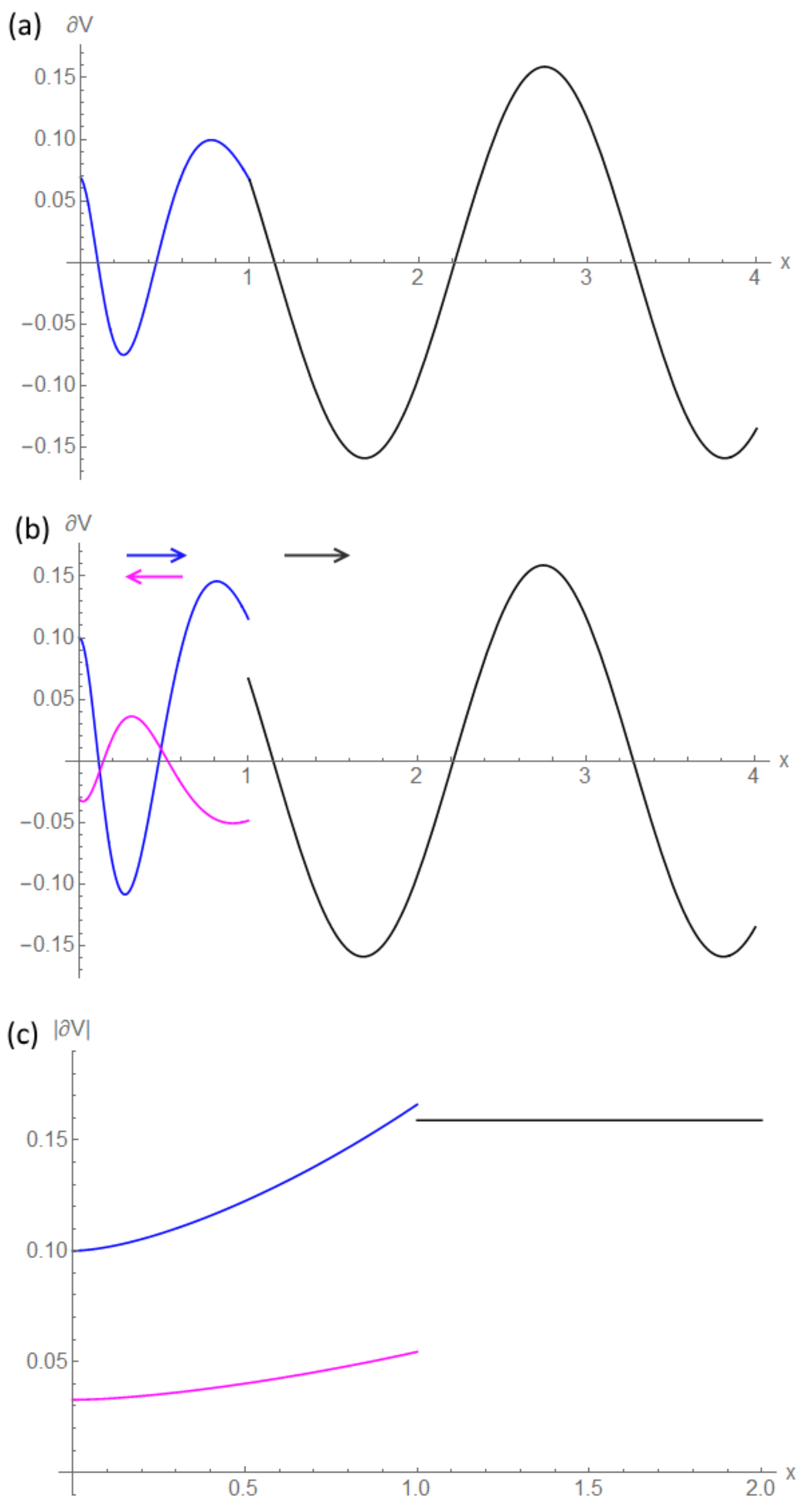}
\caption{ (a): Profile of $\delta V$ obtained from the solution of the wave equation. (b): Profiles of outgoing (blue curve), reflected (magenta) and transmitted (black) waves; (c) Profiles of amplitudes of outgoing (blue curve), reflected (magenta) and transmitted (black) waves.
\label{fig4}}
\end{figure}

Figure \ref{fig4} shows a solution of wave equation (\ref{eq3}) with boundary
conditions (\ref{bc}), background plasma profiles shown in Figure
\ref{fig2} for the parameters $\beta=0.07$ and $C_2=0.01$, arbitrary wave
amplitude $V_b=0.1$ at $x=0$ and wave angular frequency $\hat{\omega}=2 \pi \hat{\nu}$ with $\hat{\nu}=0.5$. The plasma and
magnetic field parameters for coronal plasma with $\beta=0.07$ are presented
in Table \ref{tbl-1}. The corresponding wave frequency is $0.04$ $s^{-1}$.
Figure \ref{fig4} (a) shows the profile of $\delta V$ in a fast mode wave
originating in the reconnection current sheet (at $x=0$) and propagating
outward through the non-uniform plasma (blue curve in the region 1: $0<x<1$)
and then in the uniform plasma (black curve in region 2: $x>1$) not
disturbed by reconnection. Figure \ref{fig4} (b) shows the decomposition of the wave
into the outgoing from the reconnection current sheet (blue curve), reflected
in region 1 due to gradients in plasma background (magenta) and transmitted
into the uniform plasma (black). The profiles show the real parts of the complex
expressions in Eqs. (\ref{out}),  (\ref{refl}) and (\ref{trnsm}). The
outgoing and transmitted waves propagate along the positive direction of
$x$-axis and the reflected wave propagates in the opposite direction towards
the current sheet.
The variations of the amplitudes of the three waves are shown in
Figure \ref{fig4} (c). In region 1 the amplitude of the outgoing wave
increases as it propagates outwards due to the decreasing density with
distance from the current sheet while the wave flux remains constant (the
amplitude of the reflected wave decreases toward the reconnection current
sheet for the same reason). The wavelength of the outgoing wave increases due
to the increase of the Alfv\'en speed in region 1. In region 2 where
background plasma parameters are constant, the solution is a plane wave with
constant amplitude and wavelength. Wave reflection causes a difference
between the amplitudes of the outgoing wave and the transmitted wave at $x=1$ (Figure \ref{fig4} (c)).
\begin{figure}
\includegraphics[scale=0.5]{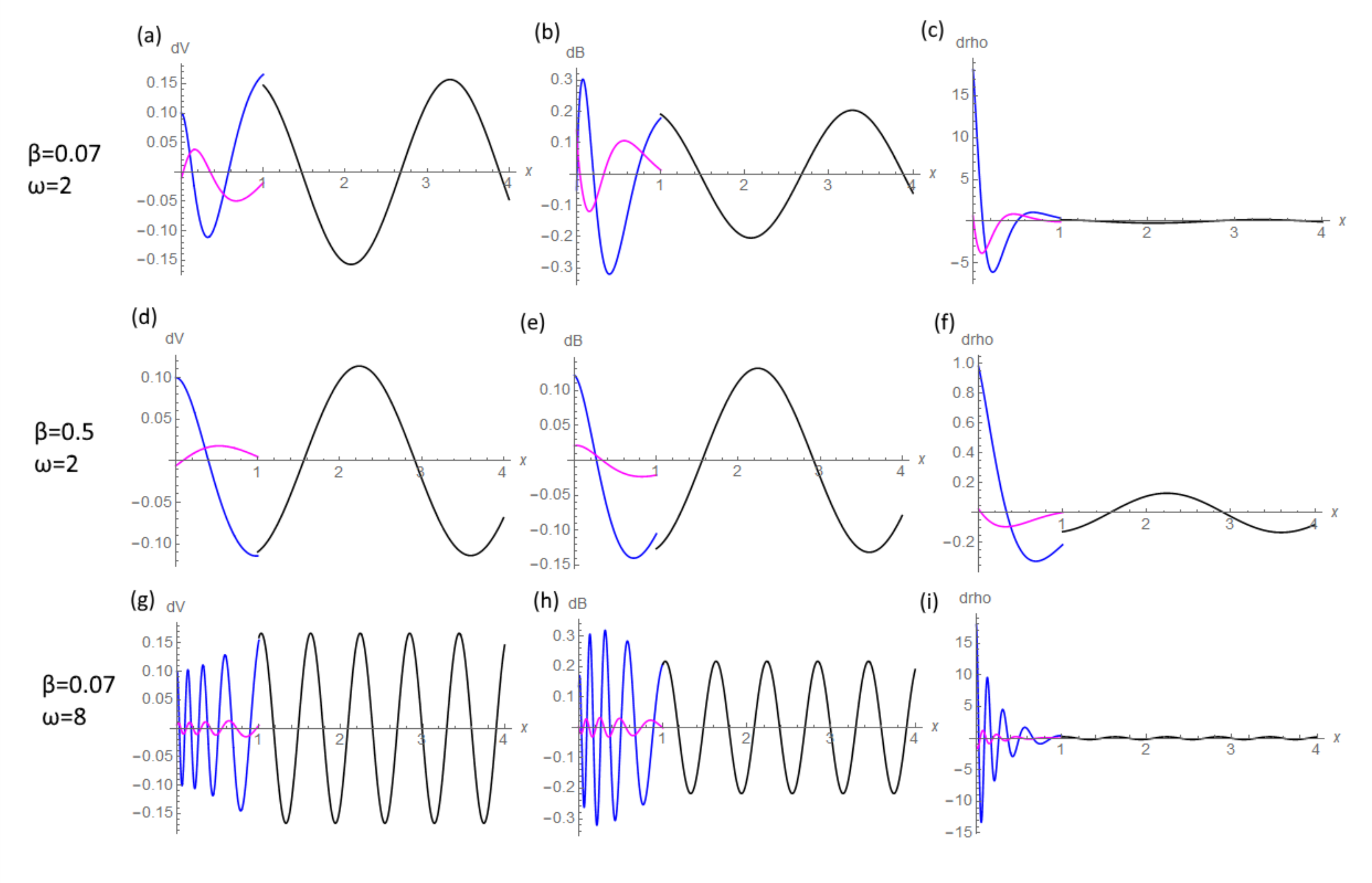}
\caption{ Profiles of velocity, density and magnetic field disturbances in outgoing (blue), reflected (magenta) and transmitted (black) fast magnetoacoustic waves for different plasma-$\beta$ and wave angular frequency: (a)-(c) $\beta=0.07$, $\hat{\omega}=2$;  (d)-(f) $\beta=0.5$, $\hat{\omega}=2$; (g)-(i) $\beta=0.07$, $\hat{\omega}=8$.
\label{fig5}}
\end{figure}

\begin{figure}
\includegraphics[scale=0.7]{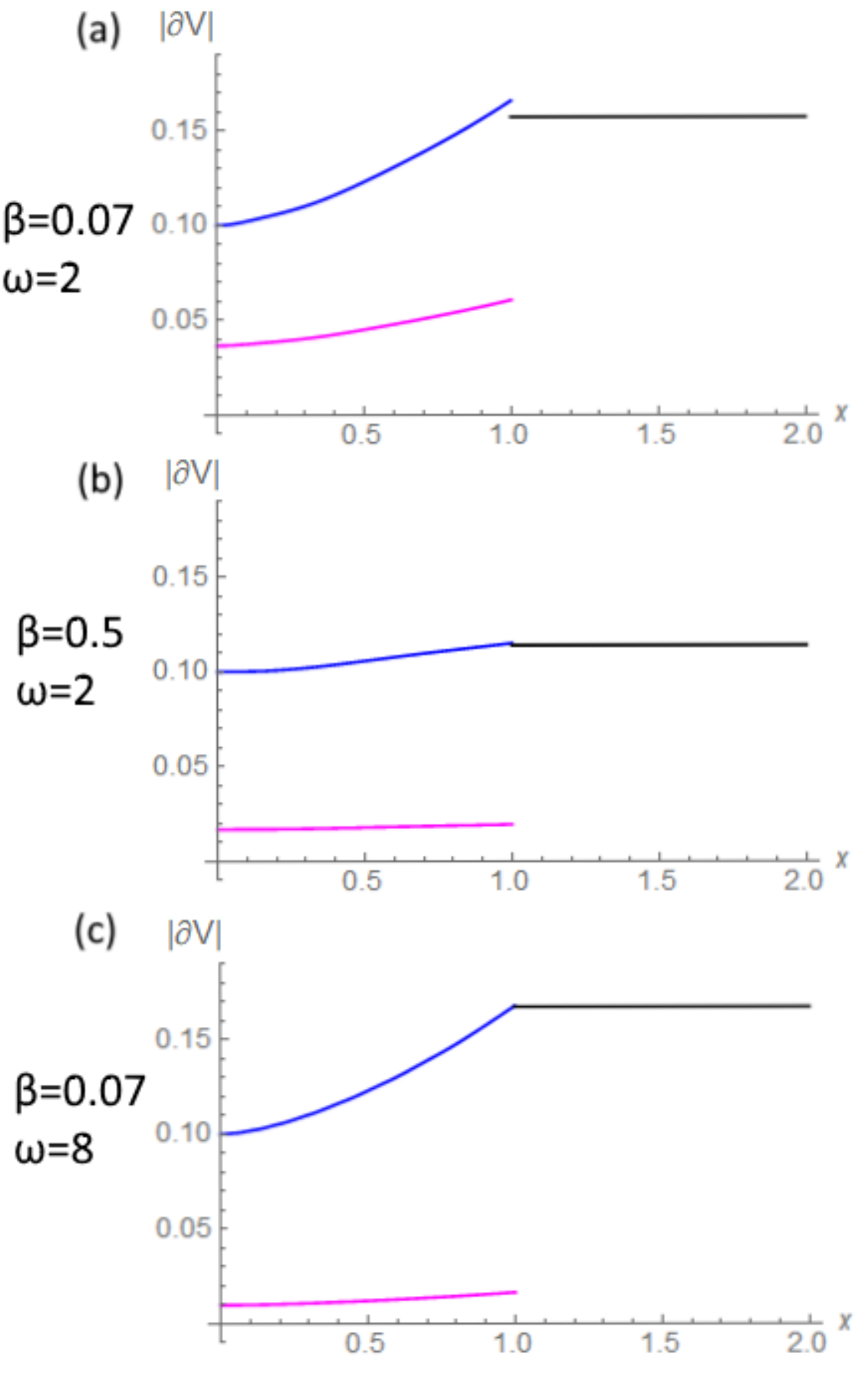}
\caption{ Amplitudes of outgoing (blue), reflected (magenta) and transmitted (black) waves for different cases; (a) $\beta=0.07$, $\hat{\omega}=2$;  (b) $\beta=0.5$, $\hat{\omega}=2$; (c) $\beta=0.07$, $\hat{\omega}=8$.
\label{fig6}}
\end{figure}

To find conditions for efficient wave reflection of fast waves near the
reconnection region we solve the wave equation for different plasma-$\beta$
and wave frequencies. Figure \ref{fig5} presents the profiles of velocity,
density and magnetic field disturbances in fast waves for plasma-$\beta$
parameters $\beta=0.07$ and $\beta=0.5$ and wave angular frequencies $\hat{\omega}=3$ and
$\hat{\omega}=8$. Figure \ref{fig6} shows the amplitude variations of $\delta V$ in
outgoing, reflected and transmitted waves in the corresponding cases. In the
lower-$\beta$ case, $\beta=0.07$, which is typical for coronal plasma, the amplitude of the reflected wave $|\delta V_r|$ is larger
compared to the higher-$\beta$ case $\beta=0.5$ demonstrating stronger wave
reflection near reconnection regions in lower-$\beta$ plasma. Also,
comparison of cases (a) and (c) in Figure \ref{fig6} with same $\beta=0.07$
but different wave angular frequencies $\hat{\omega}=2$ and $\hat{\omega}=8$ shows that the
amplitude of the reflected wave is larger for lower-frequency waves. This
suggests that the lower frequency waves produced in reconnection are
reflected more efficiently than higher frequency waves. In the higher
frequency case (Figure \ref{fig6} c)) the amplitude of the transmitted wave almost
matches the amplitude of the outgoing wave at $x=1$, meaning that the outgoing
fast wave propagates away from the reconnection region almost without reflection.

\begin{figure}[t]
\includegraphics[scale=0.36]{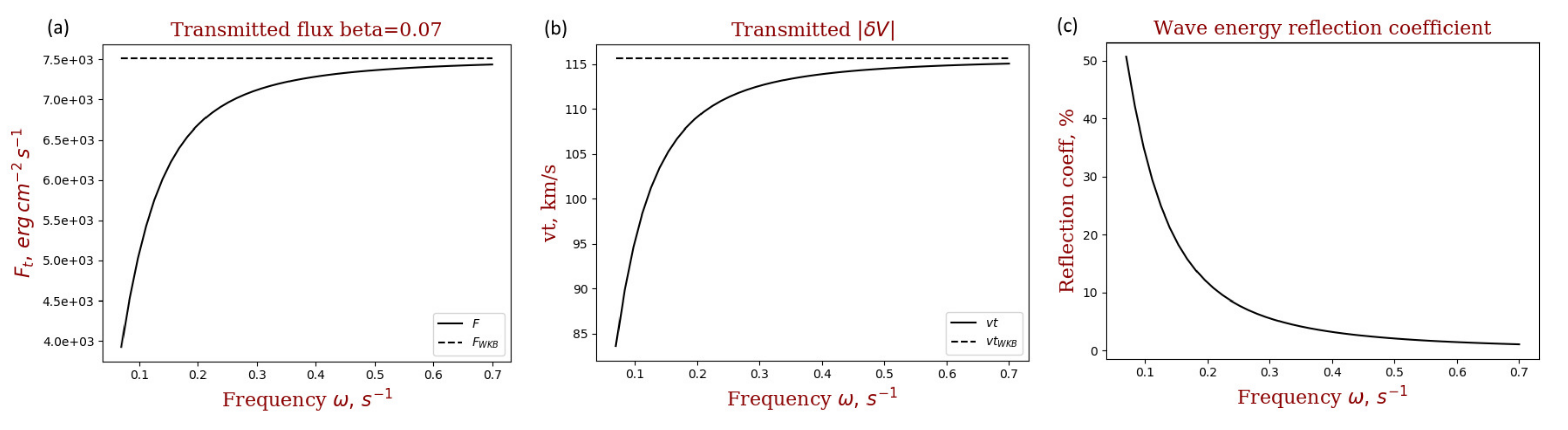}
\caption{ (a) Transmitted wave energy flux $F_t$ as a function of wave angular frequency in a range $0.07\, s^{-1}<\omega < 0.7\, s^{-1}$ and comparison with WKB flux. For all frequencies $|\delta V|_{x=0}=0.1$. (b) Amplitude of velocity disturbance in transmitted fast magnetoacoustic wave $v_t$ and comparison with WKB value; (c) Dependence of wave energy reflection coefficient on wave angular frequency.
\label{fig7}}
\end{figure}

Figure \ref{fig7} a) shows the transmitted wave energy flux $F_t$ calculated
as a sum of Poynting and acoustic fluxes according to (\ref{flux1}) as a
function of wave angular frequency in a range $0.07\, s^{-1}<\omega < 0.7\, s^{-1}$
for plasma beta $\beta=0.07$ and the comparison with the wave flux in WKB
approximation. If the WKB approximation were valid, no reflection would occur
and we would obtain the value of the flux $F_{WKB}=\rho_0(0) V_b^2 c_f (0)/2$
equal to the flux of outgoing waves at $x=0$ independent of the frequency.
This value is indicated by the horizontal dashed line marked WKB. The
calculated values of transmitted flux are lower than the WKB value because of
the wave reflection near the reconnection current sheet. The difference is
more noticeable for lower frequencies when wave reflection is more efficient.
The transmitted flux approaches the WKB value at high frequencies as the WKB
approximation becomes valid. Figure \ref{fig7} b) shows the amplitude of the
velocity disturbance in the transmitted wave $v_t$ as a function of
frequency. If the WKB approximation were valid we would obtain the value
$v_{t,WKB}=\sqrt{2 F_{WKB} /(c_{f,II} \rho_{0,II})}$ assuming that the wave
flux $F_{WKB}$ is constant in this limit. The WKB amplitude $v_t$ is shown by
the dashed horizontal line. Again, wave reflection causes lower amplitude
$v_t$ compared to the  WKB value; this effect is more noticable at lower
frequencies. Figure \ref{fig7} c) similarly shows the wave energy reflection
coefficient as a function of angular frequency as calculated according  to Eq.
(\ref{refcoef}). For lower wave frequencies $\nu \sim 0.01\, s^{-1}$, a
significant fraction of wave energy can be reflected, up to 50 \%. The wave
energy reflection coefficient drops quickly with increasing wave frequency.
Higher frequency waves propagate to the surrounding plasma almost without
reflection.

\begin{figure}[t]
\includegraphics[scale=0.7]{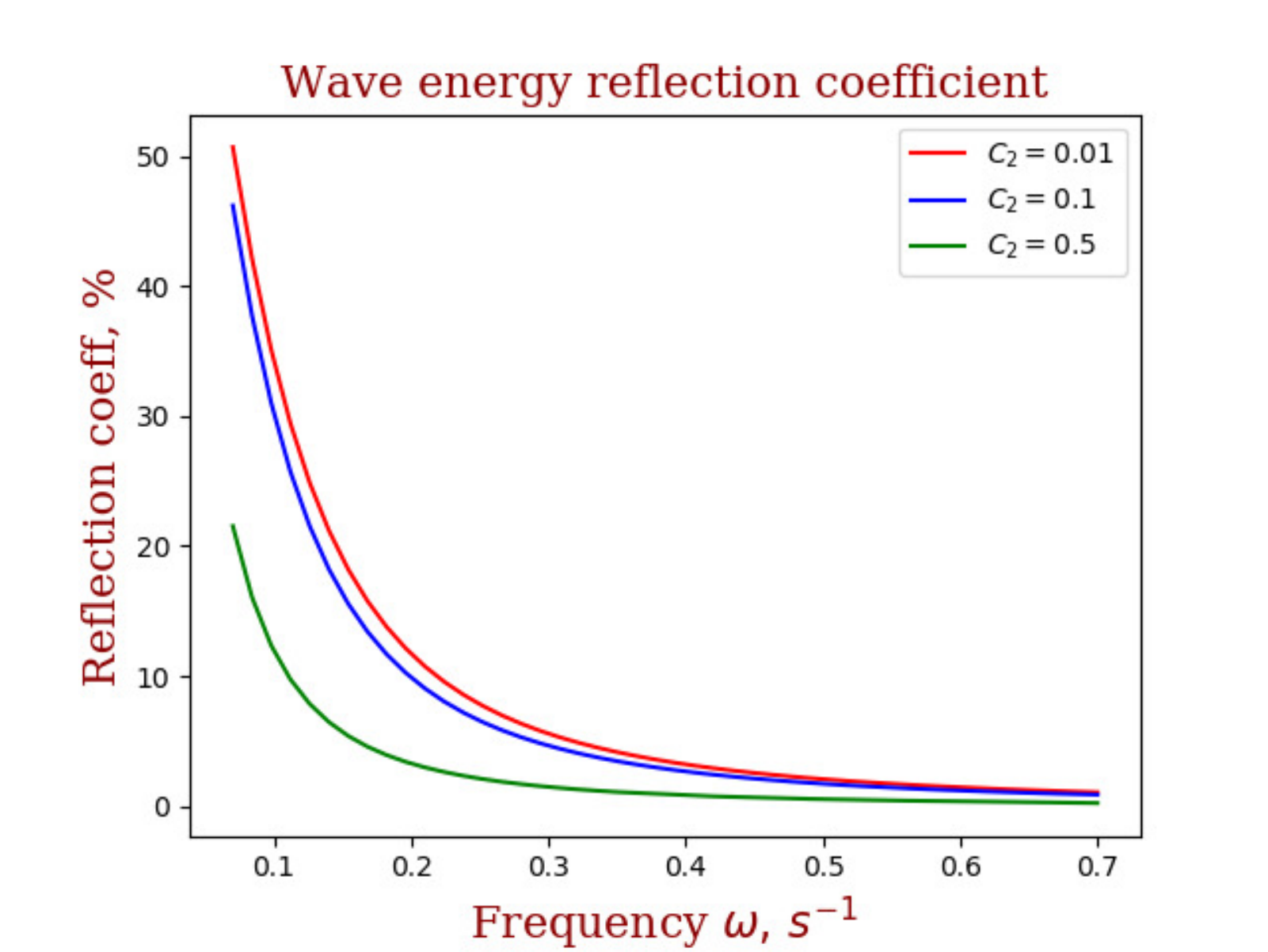}
\caption{ Dependence of wave energy reflection coefficient on wave angular frequency for different values of $C_2$. Plasma-$\beta$ parameter is $\beta=0.07$.
\label{fig8}}
\end{figure}

We have also calculated wave energy reflection coefficients for different values
of half-width of plasmoid-dominated current sheet $C_2$ (see Figure \ref{fig8}).
For the same background plasma parameters, greater $C_2$ values correspond to reconnection sites with more turbulent current sheets, where the laminar region of applicability of our analytical model is reduced.  In such reconnection sites, for a given wave frequency, a smaller fraction of wave energy is reflected back towards the current sheet.  However, one might also anticipate a broader range of wave frequencies and a higher intensity of wave generation within more turbulent reconnection current sheets.

\begin{figure}[h]
\includegraphics[scale=0.7]{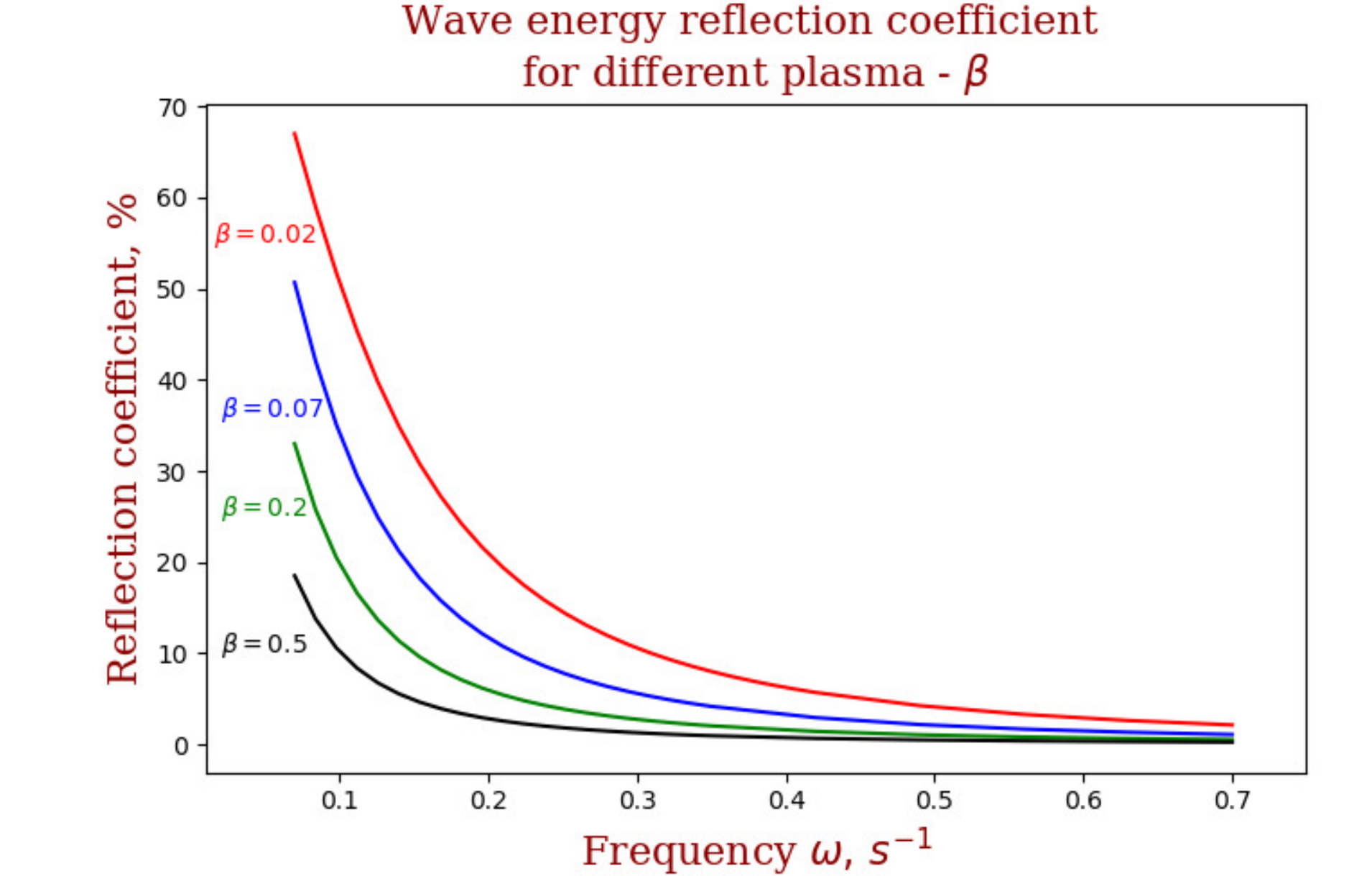}
\caption{ Wave energy reflection coefficient for different plasma-$\beta$ for a range of wave angular frequencies
\label{fig9}}
\end{figure}

The wave energy reflection coefficient $R$ as a function of wave angular frequency in the
range $0.07\, s^{-1}<\omega < 0.7\, s^{-1}$ calculated for different
plasma-beta's $\beta=0.02,\,0.07,\,0.2,\,0.5$ is shown in Figure \ref{fig9}.
The reflection coefficient is greater for waves of lower frequencies near
reconnection regions in strongly magnetized plasma with $\beta <<1$. For
example, up to 40 \% of the energy carried by waves with frequencies 0.01
$s^{-1}$ will be reflected near the reconnection region in plasma with
$\beta=0.07$ and even stronger reflection, up to 60 \%,  is expected for
 $\beta=0.02$.

\section{Role of wave reflection for particle acceleration in reconnection}\label{reflectionforparticle}

In resistive MHD simulations in \citet{prov16}, we showed that strong plasma
density compressions can form in reconnection current sheets with a guide field $B_g=0$ in
low-$\beta$ plasma. The compression is higher when the background plasma-$\beta$ is smaller, due to the extra thermal pressure required for
pressure balance with the external magnetic field. The presence of a guide
field reduces the compression, as expected. Recent kinetic simulations by
\citet{Li18} of plasmoid reconnection in low-$\beta$ plasma also show
regions of high plasma compression, although compressions appear in contracting islands rather than in current sheets. In analysis of type III radio bursts \citet{chen18}
reconstructed the trajectories of electron beams in a flare and showed the
presence of steep density gradients (5-29 Mm scale length as an upper limit)
near the flare site (see also Chen et al. (2013)). These observations support
the conclusion that reconnection regions with high compressions exist in the
corona. Results in this paper suggest that waves generated within magnetic reconnection sites  with high degree of plasma compression can undergo internal reflections due to the strong
Alfv\'en speed gradients near the reconnection region. In a lower-$\beta$
plasma the reflection becomes more efficient. In our analytical treatment we
assumed the guide field component to be zero. The effect of the guide field
would reduce the Alfv\'en speed gradient near the reconnection region and
therefore the wave reflection.

Strong plasma compression and efficient wave reflection in reconnection
regions in low-$\beta$ plasma provide appropriate conditions for Fermi
particle acceleration across the current sheet with resulting hard energy
spectra. \citet{drury12} showed that with a higher compression ratio a harder
particle energy spectrum forms (low $\gamma<5$ in a particle distribution
function $f(v)\sim v^{-\gamma}$). In this process particles bounce between
the incoming reconnecting flows scattered by waves. The presence of waves
around the reconnection region is critical for this mechanism to work so that
particles bounce multiple times across the current sheet and gain more
energy. Waves produced by reconnection and their reflection facilitate
efficient particle scattering. Further research is needed to explore the
formation of power-law distributions and confirm the relation between the
spectral index and compression ratio \citep{drury12}  in such reconnection
regions, in particular of ions, since electrons, with much smaller gyroradii,
do not participate effectively in first order Fermi acceleration across the
current sheet. Instead, electrons can be energized by a first order Fermi
acceleration process while reflecting back and forth within in the
contracting plasmoids (Drake et al. 2006).

The analytical conclusion that reconnection regions with strong compression
are a source for hard power-law ion distribution \citep{drury12} is supported
by the analysis of Fermi acceleration of electrons in plasmoids with the
incorporation of compressibility effects by \citet{montag17}. They generalized the incompressible theory of electron
acceleration in contracting plasmoids by Drake et al. (2013) and derived
analytically the dependence of spectral index $\gamma$ of electron
distribution on the compressibility $\partial n/\partial t$. Compressional
effects cause a decrease of the $\gamma$ value implying that Fermi
acceleration in the presence of compressions produces harder power-law
spectra \citep[see
also][]{Li18, leroux15}. They also found that a guide field of the order of the reconnecting
field effectively suppresses the development of a power-law distribution.

Magnetic nulls in the solar corona are thought to be key structures for
magnetic reconnection to occur. The nulls are associated with strong
gradients of magnetic field, and additionally high plasma compression can
form \citep{prov16} producing gradients of Alfv\'en speed. With these
properties magnetic nulls represent regions where magnetic reconnection can
effectively produce hard power-law distributions of particles accelerated by
first Fermi mechanism in plasmoids (electrons) and current sheets (ions).

\section{Role of reconnection generated waves for the FIP effect}\label{fip}
Alfv\'en and fast mode waves are a key agent in the chromospheric
fractionation of plasma to produce the First Ionization Potential (FIP)
Effect. \citet{pottasch63} first suggested that the elemental composition of
the solar corona might be different to that of the underlying photosphere.
Elements with FIP below about 10 eV, i.e. those like Fe, Si, and Mg, which
are predominantly ionized in the chromosphere, are seen to be enhanced in
abundance in the corona by a factor of about three. High FIP elements (e.g.
H, O, Ar) remain unchanged, though the highest FIP elements (He, Ne) may be
still further depleted. A compelling explanation for this abundance anomaly
has emerged \citep{laming15} whereby the ponderomotive force due to Alfv\'en
(or fast mode) waves propagating through or reflecting from the chromosphere
acts on chromospheric ions and in solar conditions, giving them an extra
acceleration upwards into the corona.

The most successful models of the abundance anomaly, including the extra
depletion of He and Ne, appear to arise when the Alfv\'en wave travel time
between one loop footpoint and the other is an integral number of Alfv\'en
wave half periods, i.e. when the loop is in resonance with the waves. Since
the FIP effect is widely observed in the solar corona, on different sized
loops with presumably different magnetic fields, the most plausible scenario
for the wave origin would be coronal \citep{rakowski12,laming17}, presumably
reconnection, in which waves resonant with the coronal loop would be a
natural consequence.

In stars of later spectral type than the Sun, the FIP effect decreases, and
eventually inverts for M dwarfs \citep[e.g.][]{wood13}. An ``Inverse FIP
Effect'' has also been observed in a solar flare plasma above a sunspot
\citep{doschek15}. In these cases the coronal Alfv\'en waves propagating down
to the chromosphere before reflecting back up into the corona must give way
to a population of similar waves coming up from below before reflecting back
down again. We argue that the decrease in the coronal wave amplitude at later
spectral type, which effectively means higher magnetic field, is most likely
due to the effects discussed in this paper, i.e. in higher ambient magnetic
fields (lower plasma-$\beta$), fewer waves emitted from the reconnection
current sheet can escape to infinity to cause FIP fractionation. More are
trapped locally, resulting in increased plasma heating and particle
acceleration.

\section{Conclusions}\label{conclusions}

Due to the highly unsteady, structured and impulsive nature of magnetic
reconnection, it is natural to expect the generation of MHD waves of
different modes when reconnection occurs in the solar corona. Several
observations of waves and oscillations associated with flares suggest their
origin in magnetic reconnection. Waves produced in reconnection potentially
play an important role in particle energization and elemental fractionation.
In particular, the presence of waves in reconnecting inflows is critical
for the development of magnetic turbulence scattering particles across current sheets in the first order Fermi
acceleration, and the presence of waves far from the current sheet can give
rise to a ponderomotive force to provide ion-neutral FIP fractionation. In
this work we aimed to explore how fast magnetoacoustic waves, presumably
generated by reconnection, propagate outward from the reconnection site and
what fraction of the wave energy flux is reflected and transmitted to the
surrounding plasma.

We obtained an analytical solution that describes the propagation of fast
waves in non-uniform plasma near the reconnection region. 
%Thus we took into account a decrease of plasma density that peaks at the current sheet and increase of reconnecting magnetic field magnitude with distance from the current sheet.
Due to the Alfv\'en speed gradient, fast waves
produced by unsteady reconnection can undergo reflection within the reconnection site. Wave
reflection is most efficient near the reconnection current sheets in strongly
magnetized plasma ($\beta << 1$) and for waves with lower
frequencies. We have calculated the wave energy reflection coefficient for
various plasma-$\beta$ and a range of wave frequencies. For example, for lower
frequency waves, in our calculations $\nu \sim 0.01\, s^{-1}$, and plasma
with $\beta= 0.07$, which is characteristic of the quiet solar corona, about
40 \% of wave energy flux is reflected back toward the reconnection region. The
fraction increases up to 60 \% in a lower-$\beta$ plasma $\beta=0.02$ characteristic of the coronal active regions. For
waves with higher frequencies, around $1\,s^{-1}$, the reflection coefficient
drops to 2 \%. 

We considered waves propagating perpendicular to the current
sheet with zero k-component along the B-field. While this assumption is
valid for waves with $k_x >> k_y$ (in 2D picture)  generated by elongated
plasmoids, waves with various k-vector components will be produced by the
highly irregular process of plasmoid formation and propagation. The effect of
wave reflection will still be present with the reflection coefficient
additionally depending on the direction of wave propagation. The propagation
and reflection of waves near two- and three-dimensional plasmoid-dominated
reconnection is a subject  of the future work that will combine analytical
and numerical approaches.

In the solar corona magnetic null points are topological structures 
considered as locations where magnetic reconnection is most probably to
occur. Results presented in this paper and our previous study \citep{prov16} suggest that reconnection at the nulls could also supply solar corona with suprathermal particles with hard energy spectra. Determining the acceleration mechanisms and magnetic structures 
where hard energy spectra of particles can be produced will help to
understand the origin of suprathermal ``seed''  particles in the corona. The suprathermal  population is required for production of solar energetic particles (SEP) in the
corona. Laming et al. (2013) have argued that a hard energy spectrum of
suprathermal ``seed'' particles is necessary for the injection into the
acceleration at CME shocks with low Mach number within a few solar radii from
the Sun. They suggested that these particles can originate in continuous
reconnection processes in the corona. We will further investigate the efficiency of the first-order Fermi acceleration in current sheets for ion energization, in particular in reconnection regions with multiple plasmoids and small scale current sheets, and . 

We also considered implications of wave reflection for element fractionation in stellar coronae. 
In active strongly magnetized coronae of later spectral type stars, waves
produced in magnetic reconnection would be trapped locally meaning that less
waves propagate from reconnection sites to infinity. Consequently, these
effects cause a diminishing of the chromospheric ponderomotive force that
provides extra acceleration to ions from the photosphere to the corona and
generates FIP fractionation in stellar coronae. We suggest that the efficient
wave reflection in reconnection processes in strongly magnetized stellar atmospheres  could be a possible 
explanation of the diminished FIP fractionation observed in later spectral
type.

\acknowledgments

E.P. was supported by the NASA LWS Jack Eddy Postdoctoral Fellowship. V.S.L. acknowledges support from the National Science Foundation. J.M.L. was supported by basic research funds of the Chief of Naval Research. This research was also supported by NASA Solar and Heliospheric Physics program. This work has benefited from the use of NASA's Astrophysics Data System.

\appendix

\section{Appendix}
\begin{figure}[h]
\includegraphics[scale=0.55]{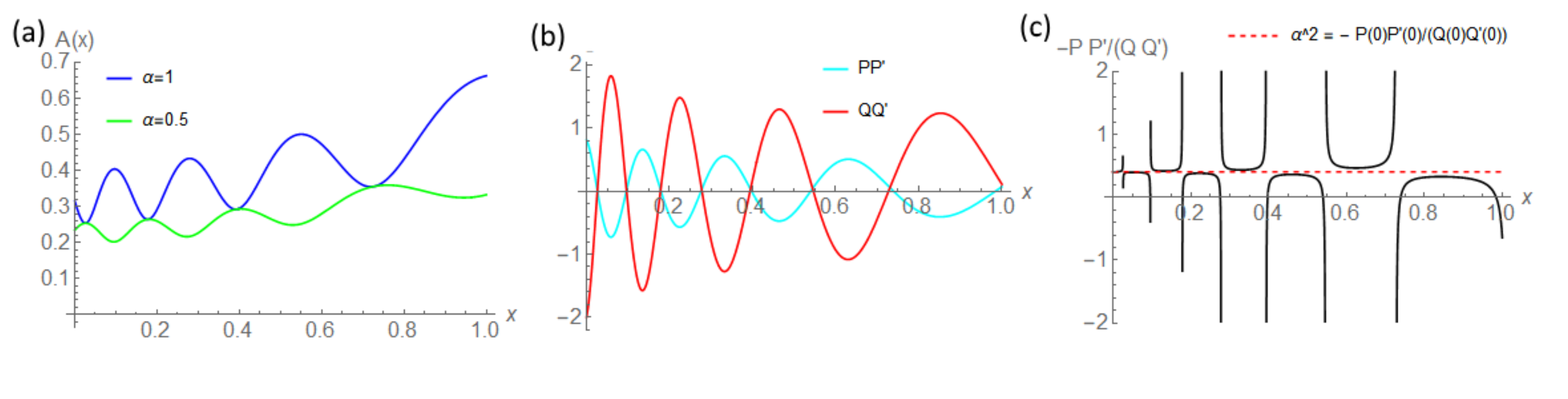}
\caption{ (a) Amplitude $A(x) = \sqrt{P_p^2 + \alpha^2 Q_p^2}$ as a function of $x$ for the order $p=9.4$ of Legendre functions and two arbitrary values of $\alpha = 0.5;\, 1$. (b) $P_p(x)P'_p(x)$ and $Q_p(x)Q'_p(x)$ as functions of $x$ for the order $p=9.4$. (c) Black curve: function $-\frac{P_p(x)P'_p(x)}{Q_p(x)Q'_p(x)}$ for the order $p=9.4$. Red dashed line: the constant $\alpha^2 = - \frac{P(0)P(0)'}{Q(0)Q(0)'}$.
\label{fig10}}
\end{figure}

The solution of equation (\ref{eqxi}) for the coordinate dependent part
$\delta V (x)$ of the velocity  disturbance $\delta {V'} (x,t) = \delta V(x)
\exp (-i \omega t)$ is a linear combination of the associated Legendre
functions of first and second kind $P_p(x)$ and $Q_p(x)$. We will look for a
linear combination that represents  a sum of counter propagating waves,
outgoing and reflected, which has a form
\begin{eqnarray}
\delta V(x) = c (P_p(x)-i \alpha Q_p(x))+d(P_p(x)+i \alpha Q_p(x))
\end{eqnarray}
For an arbitrary value of $\alpha$, the amplitudes of these two waves,
$A_{out}(x)=|c| \sqrt{P_p^2 + \alpha^2 Q_p^2}$ and $A_r(x)=|d| \sqrt{P_p^2 +
\alpha^2 Q_p^2}$ respectively, are oscillating functions of $x$ (see Fig
\ref{fig10} (a)). That means that with arbitrary $\alpha$ each of the two
waves contains an outgoing and a reflected component. However a value of
$\alpha$ exists when $A(x) = \sqrt{P_p^2 + \alpha^2 Q_p^2}$ is a
monotonically increasing function of $x$, e.g. for this $\alpha$, $A'(x) \geq
0$. In this case the combinations $c (P_p(x)-i \alpha Q_p(x))$ and
$d(P_p(x)+i \alpha Q_p(x))$  define purely outgoing and reflected waves,
respectively. To find $\alpha$ we solve inequality $A'(x)=P_p P'_p + \alpha^2
Q_p Q'_p  \geq 0$. When $Q_p Q'_p=0$ the product $P_p P'_p$ also equals 0
(see Fig \ref{fig10} (b)) so the inequality holds. At all intervals where $Q_p
Q'_p > 0$ (Fig \ref{fig10} (b)) the condition for the $\alpha$ parameter is
$\alpha^2 \geq -\frac{P_p(x)P'_p(x)}{Q_p(x)Q'_p(x)}$. The function
$-\frac{P_p(x)P'_p(x)}{Q_p(x)Q'_p(x)}$ is shown in Fig \ref{fig10} (c). For
intervals where $Q_p Q'_p < 0$ (Fig 1A (b)) the condition for $\alpha$ is
$\alpha^2 \leq -\frac{P_p(x)P'_p(x)}{Q_p(x)Q'_p(x)}$. The only value of
$\alpha$ that satisfies the two conditions at all intervals is $\alpha =
\sqrt{- \frac{P_p(0)P'_p(0)}{Q_p(0)Q'_p(0)}}$. Figure \ref{fig10} (c) shows
the value of $\alpha^2 = - \frac{P_p(0)P'_p(0)}{Q_p(0)Q'_p(0)}$ by the red
dashed  line. With this choice of the $\alpha$ parameter, the amplitudes of
the outgoing and reflected waves are monotonically increasing functions as
shown in Figures \ref{fig4} c) and \ref{fig6}.

\clearpage

\end{document}